\documentclass[letter]{aa}
\usepackage{graphicx}
\usepackage{txfonts}

\begin{document}

\title{GRB 050223: A dark GRB in a dusty starburst galaxy\thanks{Based
on observations collected at the European Southern\- Ob\-ser\-va\-to\-ry,
Chile, under proposal ESO~75.A-0718.}}

\author{L.~J. Pellizza\inst{1,2} \and P.-A. Duc\inst{1} \and E. Le
Floc'h\inst{3,4} \and I.~F. Mirabel\inst{5} \and L.~A. Antonelli\inst{6} \and
S. Campana\inst{7} \and G. Chincarini\inst{7,8} \and A. Cimatti\inst{9} \and S.
Covino\inst{7} \and M. Della Valle\inst{9} \and F. Fiore\inst{6} \and D.
Fugazza\inst{7} \and P. Giommi\inst{10} \and P. Goldoni\inst{11} \and G.~L.
Israel\inst{6} \and E. Molinari\inst{7} \and A. Moretti\inst{7} \and L.
Piro\inst{12} \and P. Saracco\inst{7} \and L. Stella\inst{6} \and G.
Tagliaferri\inst{7} \and M. Vietri\inst{13}}

\institute{Laboratoire AIM (UMR 7158), Service d'Astrophysique, CEA Saclay,
L'Orme des Merisiers, B\^at. 709, 91191 Gif-sur-Yvette, France \and Instituto
de Astronom\'{\i}a y F\'{\i}sica del Espacio (CONICET/UBA) C.C. 67, Suc. 28,
1428 Buenos Aires, Argentina \and Steward Observatory, University of Arizona,
933 North Cherry Avenue, Tucson, AZ 85721, USA \and Chercheur Associ\'e,
Observatoire de Paris, F-75014, Paris, France \and European Southern
Observatory, Alonso de C\'ordova 3107, Santiago 19, Chile (on leave from CEA
Saclay, France) \and INAF, Osservatorio Astronomico di Roma, via di Frascati
33, I-00040 Monteporzio Catone (Roma), Italy \and INAF, Osservatorio
Astronomico di Brera, via E. Bianchi 46, I-23807 Merate (LC), Italy \and
Universit\`a degli studi di Milano-Bicocca, Dipartimento di Fisica, piazza
delle Scienze 3, I-20126 Milano, Italy \and INAF, Osservatorio Astrofisico di
Arcetri, largo E. Fermi 5, I-50125 Firenze, Italy \and Agenzia Spaziale
Italiana, Science Data Center, via di Villa Grazioli, I-00198 Roma, Italy \and
Laboratoire Astroparticule et Cosmologie, UMR 7164, 11 Place Marcelin
Berthelot, 75231 Paris Cedex 05, France \and INAF-IASF, Sezione di Roma, via
Fosso del Cavaliere 100, I-00133 Roma, Italy \and Scuola Normale Superiore,
piazza dei Cavalieri 7, I-56126 Pisa, Italy}

\offprints{L.~J. Pellizza, \email{pellizza@iafe.uba.ar}}

\date{Received / Accepted}

\abstract{}{We aim at detecting and determining the properties of the host
galaxy of the dark GRB 050223.}{We use VLT optical/NIR images coupled
to {{\em Swift}} X-ray positioning, and optical spectra of the host galaxy to
measure its properties.}{We find a single galaxy within the {{\em Swift}} error
box of GRB 050223. It is located at $z = 0.584$ and its luminosity is
$L \sim 0.4$~$L^*$. Emission lines in the galaxy spectrum imply an
intrinsic
${\rm SFR} > 7$~$M_{\sun}$~yr$^{-1}$, and a large extinction $A_{\rm V} >
2$~mag within it. We also detect absorption lines, which reveal an underlying
stellar population with an age between 40~Myr and 1.5~Gyr.}{The identification
of a host galaxy with atypical properties using only the X-ray transient
suggests that a bias may be present~in the former sample of host galaxies. Dust
obscuration together with intrinsic faintness are the most probable causes for
the darkness of this burst.}

\maketitle

\section{Introduction}
\label{intro}

Long gamma-ray bursts (LGRBs) are brief pulses of $\gamma$-ray radiation
occuring about once per day at random positions in the sky, and lasting more
than 2~s. LGRBs are accompanied by fading afterglows detected at longer
wavelengths, which can be observed for months or even years. Optical and NIR
afterglows are usually observed, except for a subpopulation of dark LGRBs
(DGRBs). This can be attributed, in some cases, to the lack of early
observations reaching faint magnitudes. In other cases deep and prompt
follow-ups indicate that true DGRBs indeed exist, and that they comprise a
fraction of $\sim$10\% of all bursts (Jakobsson et~al. \cite{Jak05}).

The lack of optical afterglows in DGRBs could be explained in different ways.
DGRB afterglows could be intrinsically underluminous, or extincted by
intergalactic Ly-$\alpha$ absorbers (if located at high redshift) or by the
dusty interstellar medium of their host galaxies (HGs). While some authors find
that the intrinsic faintness hypothesis could explain some cases (e.g.,
Pedersen et~al. \cite{Ped06}), for other DGRBs the existence of at least some
obscuration inside the HG has been claimed (e.g., Piro et~al. \cite{Pir02}). A
particularly interesting case is GRB 970828 (Groot et~al.
\cite{Gro98}; Djorgovski et~al. \cite{Djo01}), which has been proposed as the
prototype of dust-enshrouded bursts, and is the only DGRB whose HG has been
detected in the FIR by {\em Spitzer} (Le Floc'h et~al. \cite{LeF06}).

The hypothesis of dust extinction of DGRB afterglows is particularly
interesting in the context of the relationship between LGRBs and star formation
(SF). The connection of LGRBs with core-collapse supernovae (Hjorth et~al.
\cite{Hjo03}; Stanek et~al. \cite{Sta03}; Malesani et~al. \cite{Mal04};
Campana et~al. \cite{Cam06}), the SF activity observed in their HGs (Le Floc'h
et~al. \cite{LeF02}; Prochaska et~al. \cite{Pro04}), and their massive stellar
progenitors proposed by theoretical models (Fryer et~al. \cite{Fry99}), suggest
that LGRBs might be good SF tracers. Their detectability at high redshift
(Tagliaferri et~al. \cite{Tag05}; Haislip et~al. \cite{Hai06}; Kawai
et~al. \cite{Kaw06}) would then transform them in a promising tool to probe SF
in the early Universe. However, the sample of HGs obtained so far shows a bias
to underluminous and blue objects (Le Floc'h et~al. \cite{LeF03}), while there
is evidence that a significant fraction of the SF at high redshift took place
in luminous, reddened and dust-enshrouded IR starbursts (Elbaz et~al.
\cite{Elb02}). If the obscuration hypothesis is correct, a natural explanation
for this bias arises from the fact that, in most previous investigations, LGRB
HGs have been identified through the accurate positioning given by optical
afterglow measurements. Hence, mostly HGs of unobscured bursts have been
identified, leading to a natural exclusion of dusty starburst galaxies. The
detection of such galaxies associated to LGRBs would make a strong case for the
obscuration hypothesis, and would also support the LGRB-SF connection.

GRB 050223 was discovered by the {\em Swift} satellite at 03:09:06~UT
on February 23, 2005 (Mitani et~al. \cite{Mit05}), and also detected by
{\em INTEGRAL} (Mereghetti et~al. \cite{Mer05}). It was a LGRB ($T_{90} =
23$~s), with a fluence of $7.4 \times 10^{-7}$~erg~cm$^{-2}$ (15--350~keV). The
{\em Swift} X-ray telescope (XRT) detected its afterglow (Giommi et~al.
\cite{Gio05}), which was also observed by {\em XMM-Newton} (De Luca \& Campana
\cite{DeL05}). The afterglow presented a power-law decay with a slope $\alpha =
0.99^{+0.15}_{-0.12}$. Its spectrum was well fitted by an absorbed power law
with a photon index $\Gamma = 1.75^{+0.19}_{-0.18}$, and a hydrogen density
column $N_{\rm H} \sim 10^{21}$~cm$^{-2}$, consistent with the Galactic value,
$N_{\rm H} = 7 \times 10^{20}$~cm$^{-2}$ (Page et~al. \cite{Pag05}). XRT gave
also a precise positioning of the afterglow, $\alpha_{\rm J2000} =
18^{\rm h}05^{\rm m}33\fs08$, $\delta_{\rm J2000} = -62\degr28\arcmin20\farcs5$
(5\farcs4 error radius; Moretti et~al. \cite{Mor06}).

Searches for the ultraviolet and optical afterglow were made by {\em Swift}
Ultraviolet and Optical Telescope (Gronwall et~al. \cite{Gro05}), {\em XMM}
Optical Monitor (Blustin et~al. \cite{Blu05}), ROTSE~III (Smith \cite{Smi05}),
PROMPT (Nysewander et~al. \cite{Nys05}) and the Mount John University
Observatory (Gorosabel et~al. \cite{Gor05}). No observations were reported in
the IR or radio domains. The deepest observations put a stringent limiting
magnitude $R > 21.2$ only 4.1 hours after the burst. Assuming that the decay of
the optical afterglow follows the same law as the X-ray afterglow, and taking
into account the Galactic absorption of 0.2~mag at the position of the burst
(Schlegel et~al. \cite{Sch98}), this limit implies a magnitude $R > 23.8$
at 2 days after the burst. Hence, following the criterion of Djorgovski et~al.
(\cite{Djo01}), GRB 050223 can be classified as a DGRB. The
aforementioned observations put also an upper limit $\beta_{\rm OX} < 0.8$ for
the mean optical to X-ray spectral index of GRB 050223, hence it can
not be classified as dark according to the criterion of Jakobsson et~al.
(\cite{Jak04}), which requires $\beta_{\rm OX} < 0.5$ for a burst to be dark.
This is a direct consequence of the faintness of the X-ray afterglow, which was
used by Page et~al. (\cite{Pag05}), together with the value of $N_{\rm H}$, to
argue that GRB 050223 has an intrinsically underluminous afterglow.

In this Letter we report the detection of the host galaxy of GRB
050223 and the determination of its properties. Sects.~\ref{obs} and \ref{res}
present our observations and results respectively. Sect.~\ref{disc} discusses
the properties of the galaxy and their implications on the nature of DGRBs.

\section{Observations}
\label{obs}

Our observations of GRB 050223 were carried out with the ESO 8-meter
Very Large Telescope UTs 1 and 2 at Paranal Observatory, Chile, as part of a
program dedicated to the study of LGRB HGs (ESO 75.A-0718, P.~I. Pellizza).
Optical and NIR images of the field of the source were obtained with the ESO
Focal Reducer and Low Dispersion Spectrograph 1 (FORS~1) on May 10, 2005, and
with the Infrared Spectrometer And Array Camera (ISAAC) on June 5, 2005. FORS~1
was used in imaging mode with the Bessel R filter to obtain 10 deep (308~s per
exposure) frames of the field of the source. A set of 42 1~minute exposures
were also taken in the short wave mode of ISAAC with the $K_{\rm s}$ filter.
These images were reduced in the standard way (bias/dark subtraction, flat
fielding and illumination correction), and combined to obtain very deep final
science images with equivalent exposures of 51 and 42~minutes in the $R$ and
$K_{\rm s}$ bands, respectively.

By chance, we found a single HG candidate within GRB 050223 {\em
Swift} error box, down to the sensitivity limits of our data. For this
candidate, we obtained optical spectra with FORS~2 on July 30--31, 2005. Three
1~h exposures were made in LSS mode with an 1{\arcsec} slit, grism 300I and
the OG590 order sorting filter. Each of them was reduced (bias subtraction,
flat fielding), and the three frames were combined into a single one with an
equivalent exposure of 3~hours, from which the spectrum was extracted and
calibrated in wavelength and flux. All reduction and calibration tasks were
performed with IRAF.

\section{Data analysis and results}
\label{res}

\begin{figure}
\resizebox{\hsize}{!}{\includegraphics{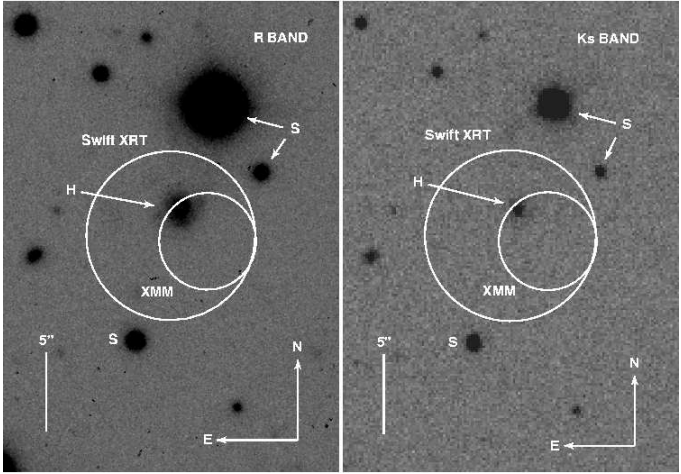}}
\caption{The field of GRB 050223 in $R$ (left panel) and $K_{\rm s}$
(right panel) bands. North is up and East is to the left. The large and small
circles are the {\em Swift} XRT and {\em XMM} error boxes for the source
(5\farcs4 and 3\farcs1 radius respectively at 90\% CL). A single extended
object (H) is found inside them in both images. Other three objects (S) lying
near their boundaries are stars with $>99\%$ probability.}
\label{img}
\end{figure}

We performed the astrometry of our science $R$ and $K_{\rm s}$ images
using 2MASS coordinates of 22 detected stars. We obtained an rms error of
0\farcs13 and 0\farcs14 respectively, small enough for our purposes. In
Fig.~\ref{img} we show the field of the source in both bands, together with the
position of the {\em Swift} and {\em XMM} error circles (5\farcs4 and 3\farcs1
radius at 90\% CL respectively), the latter obtained from our own analysis of
{\em XMM} data. By chance, we find a single object (H in Fig. \ref{img}) inside
both circles. Three other objects (S in Fig.~\ref{img}) lie outside them but
near their boundaries; they are stars with $>$99\% probability according to a
PSF fit to their profiles. Object H is extended (roughly 1\farcs5 in radius)
and morphologically consistent with a galaxy. Its coordinates (J2000) are
$\alpha = 18^{\rm h}05^{\rm m}32\fs992$, $\delta =
-62\degr28\arcmin18\farcs81$ ($\pm$0\farcs16, which takes into account the
2MASS catalog uncertainty of 0\farcs1), only 1\farcs8 apart from the
nominal XRT afterglow position, and 2\farcs7 from that of {\em XMM}. There is
no other conspicuous object within the error circles down to our $3\sigma$
limiting magnitudes of $R\sim26$ and $K_{\rm s}\sim21$. We note that with these
limits a typical LGRB HG like those in the sample of Le Floc'h et~al.
(\cite{LeF03}) would be detectable up to a redshift of $\sim$2.5. Hence,
although at this point we cannot completely discard a high redshift galaxy, H
is a good HG candidate for GRB 050223.

We performed differential aperture photometry of candidate H, using a set of
stars in our frames with known USNO~B1.0 and 2MASS as comparisons. $R$
magnitudes of the comparison stars were computed from their photographic
IIIa-J and IIIa-F magnitudes and the transformations of Windhorst et~al.
(\cite{Win91}), while their $K_{\rm s}$ magnitudes were taken directly from the
2MASS catalog. Linear fits of comparison against instrumental magnitudes give
$R = 21.76 \pm 0.05$ and $K_{\rm s} = 18.88 \pm 0.02$ for candidate H.
Correcting for Galactic extinction ($E(B-V) = 0.09$; Schlegel et ~al.
\cite{Sch98}), we obtain its intrinsic magnitudes, $R = 21.55 \pm 0.05$ and
$K_{\rm s} = 18.85 \pm 0.02$. Using galaxy number counts (Liske et~al.
\cite{Lis03}), we derive probabilities of $\sim$5\% and $\sim$2\% of finding a
galaxy brighter than candidate H within the XRT and {\em XMM} error circles,
respectively. Computing the fraction of the sky area covered by galaxies
brighter than H (e.g., Piro et~al. \cite{Pir02}) gives a probability of chance
alignment of 1\%, consistent with the previous values. Hence, we can discard
at $>$95\% CL the hypothesis that H is a field galaxy aligned by chance
with the burst, making a stronger case for its association to GRB
050223.

\begin{figure}
\resizebox{\hsize}{!}{\includegraphics{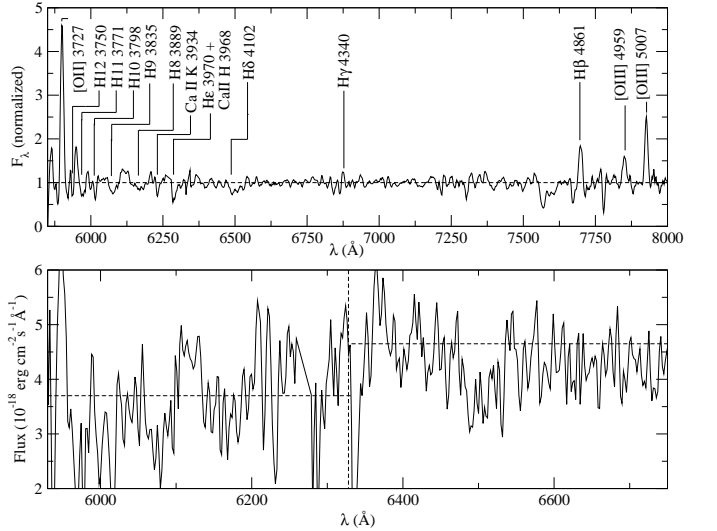}}
\caption{{\em Upper panel:} Normalized optical spectrum of candidate H, showing
lines both in emission (H$\beta$, H$\gamma$, [\ion{O}{ii}], [\ion{O}{iii}]) and
absorption (H$\delta$, \ion{Ca}{ii}, early Balmer lines). {\em Lower panel:}
Flux-calibrated spectrum of the region around 4000~{\AA} (rest frame, vertical
dashed line), exhibiting the break at this wavelength. Horizontal dashed lines
are plotted at the mean flux in the 3750--3950~{\AA} and 4050--4250~{\AA}
ranges.}
\label{spec}
\end{figure}

In Fig.~\ref{spec} we show our spectrum of candidate H; the features identified
in it are listed in Table~\ref{lines}. The spectrum presents prominent emission
lines (H$\beta$, H$\gamma$, [\ion{O}{ii}] 3727~{\AA}, [\ion{O}{iii}]
4959,5007~{\AA}). Although the S/N ratio is poor, we identify also a rather
strong stellar continuum exhibiting a clear break at 4000{\AA} (rest frame),
and absorption lines such as strong H$\delta$, \ion{Ca}{ii} lines, and several
early Balmer lines. From the emission lines we derive a redshift $z = 0.584 \pm
0.005$, consistent with that obtained by Berger \& Shin (\cite{Ber06}). At this
redshift, the luminosity distance of the galaxy (for a cosmological model with
$\Omega_\Lambda = 0.7$, $\Omega_{\rm m} = 0.3$ and $H_0 =
70$~km~s$^{-1}$~Mpc$^{-1}$) is 3.4~Gpc. Using this distance and the
k-correction for an Sc model (Poggianti \cite{Pog97}), we derive an absolute
magnitude $M_{K_{\rm s}} = -23.47 \pm 0.03$. Assuming $M^*_{K_{\rm s}} = -24.5$
(e.g., Saracco et~al. \cite{Sar06}) the luminosity of the galaxy is
$L\sim$0.4~$L^*$, which makes H one of the most luminous LGRB
HGs. The blue $R - K_{\rm s} = 2.70 \pm 0.07$, though not unusual, is redder
than those of LGRB HGs at $z \sim 0.6$ (Le Floc'h et~al. \cite{LeF03}).

\begin{table}
\begin{center}
\caption{Lines identified in the spectrum of candidate H.}
\label{lines}
\begin{tabular}{lccc}
\hline
\hline
Species & $\lambda_{\rm lab}$ & $\lambda_{\rm obs}$ & EW (rest frame) \\
 & ({\AA}) & ($\pm1$~{\AA}) & ({\AA}) \\
\hline
[\ion{O}{ii}] & 3727 & 5900 & $-27 \pm 6$ \\
H12 & 3750.154 & 5937 & $2.3 \pm 0.3$ \\
H11 & 3770.632 & 5972 & -- \\
H10 & 3797.900 & 6015 & $2.6 \pm 0.6$ \\
H9 & 3835.386 & 6077 & -- \\
H8 & 3889.051 & 6167 & -- \\
\ion{Ca}{ii} K & 3933.664 & 6232 & $<2.6$ \\
\ion{Ca}{ii} H & 3968.470 & 6285 & -- \\
H$\varepsilon$ & 3970.074 & 6289 & -- \\
H$\delta$ & 4101.737 & 6496 & $4.4 \pm 0.6$ \\
H$\gamma$ & 4340.468 & 6875 & $>-2.4$ \\
H$\beta$ & 4861.332 & 7699 & $-8.2 \pm 2.0$ \\
$[\ion{O}{iii}]$ & 4959 & 7850 & $-5.7 \pm 1.8$ \\
$[\ion{O}{iii}]$ & 5007 & 7927 & $-11.3 \pm 1.3$ \\
\hline
\end{tabular}
\end{center}
\end{table}

The detection of H$\delta$ and early Balmer lines in absorption, while
H$\gamma$ and H$\beta$ are seen in emission, implies a strong Balmer decrement
and suggests a large extinction. Assuming $R_V = 3.1$, the ratio ${\rm
H}\gamma / {\rm H}\beta < 0.3$ (corrected for Galactic extinction) gives
$A_{\rm V} > 2$~mag, or $E(B-V) > 0.65$ within H. This corresponds to a
hydrogen column density $N_{\rm H} \gtrsim 10^{21}$~cm$^{-2}$ (Dickey \&
Lockman \cite{Dic90}), marginally consistent with that derived from X-ray
observations (Page et~al. \cite{Pag05}). The observed emission lines indicate
the presence of ongoing star formation. We measure a lower limit\footnote{The
galaxy is larger than the slit width, but we expect this limit to be near the
true value because only the faint outskirts were missed.} to the [\ion{O}{ii}]
3727~{\AA} flux $F_{[\ion{O}{ii}]} > 8.2 \times
10^{-17}$~erg~cm$^{-2}$~s$^{-1}$, corrected for Galactic extinction, which
corresponds to a luminosity $L_{[\ion{O}{ii}]} > 1.3 \times
10^{41}$~erg~s$^{-1}$. Correcting for the measured extinction within H,
we derive a ${\rm SFR} > 7$~$M_{\sun}$~yr$^{-1}$ (Kennicutt \cite{Ken98}), and
hence a specific SFR $>17.5$~$M_{\sun}$~yr$^{-1}$~$L^{*-1}$. We do
not detect the [\ion{Ne}{iii}] 3869~{\AA} line, implying a
low~[\ion{Ne}{iii}]/[\ion{O}{ii}]~ratio.

The rest-frame equivalent widths of H$\delta$ and [\ion{O}{ii}]
3727~{\AA} are consistent with an e(a) classification for this galaxy ---e(a)
galaxies are emission-line objects with strong H$\delta$ absorption, ${\rm
EW}({\rm H}\delta) > 4$~{\AA}, and moderate [\ion{O}{ii}] emission,
5~{\AA}~$<|{\rm EW}(\ion{O}{ii})|< 40$~{\AA} (Poggianti \& Wu
\cite{Pog00})---.
This is
also consistent with the derived extinction, as e(a) galaxies have a median
$E(B-V) = 1.11$. The probability of finding an e(a) galaxy within the XRT error
circle is less than 0.5\% (Poggianti \& Wu \cite{Pog00}), making a stronger
case for the association between this galaxy and GRB 050223. The
presence of a strong H$\delta$ line and several early Balmer lines
(H$\varepsilon$--H12) in absorption indicates also the existence of a stellar
population dominated by A0V stars, whose age could range from 40~Myr
to 1.5 Gyr (Poggianti \& Wu \cite{Pog00}). Finally, the observed 4000~{\AA}
break has an index $D_{4000} = 1.3 \pm 0.2$, which is consistent with late-type
star-forming galaxies.

\section{Discussion}
\label{disc}

Combining the accurate X-ray positioning of LGRB afterglows given by {\em
Swift} XRT, and very deep VLT optical and NIR images and optical spectra, we
identified the likely HG of the dark GRB 050223. It is a
0.4~$L^*$,
blue galaxy at $z = 0.584$, with strong SF (${\rm SFR} >
7$~$M_{\sun}$~yr$^{-1}$, SSFR~$>
17.5$~$M_{\sun}$~yr$^{-1}$~$L^{*-1}$), a stellar population
dominated
by A0V stars with an age of 40~Myr--1.5~Gyr, and a large
extinction ($A_{\rm V} > 2$~mag).

This galaxy has some typical properties of LGRB HGs. It is underluminous and
blue, with strong emission lines indicating SF, although it is brighter and
redder than other LGRB HGs at its redshift (Le Floc'h et~al. \cite{LeF03}). It
shows also some atypical features. The [\ion{Ne}{iii}] 3869~{\AA} to
[\ion{O}{ii}] 3727~{\AA} ratio is low, and the inferred SFR and extinction are
among the highest ones derived from optical observations (much higher SFRs were
determined from radio observations, e.g. by Berger et~al. (\cite{Ber03}),
although radio SFRs are in some cases higher than optical ones for the same
galaxy). Another atypical property is the presence of a strong stellar
population dominated by A0V stars, implied by the observed absorption lines,
and the e(a) classification that follows from them. We point out that our
finding of an HG spectrum that differs from those previously reported is the
result of a new HG selection based only on X-ray transients. This may imply the
presence of a bias in the HG sample that has been considered so far.

The large observed extinction suggests that this DGRB took place in a
dust-enshrouded starburst. The e(a) type spectrum is usually interpreted as
produced in galaxies with strong ongoing star formation, together with older
stellar populations and a high selective obscuration affecting the youngest
stars (Poggianti et~al. \cite{Pog01}). Star forming regions, inside which LGRBs
occur, are expected to be highly obscured in these galaxies. Hence, our
results support the obscuration hypothesis for DGRBs, which was also
strengthened by the recent classification of the HG of GRB 030115 as
an extremely red object (Levan et~al. \cite{Lew06}). Our picture is consistent
with that of Nakagawa et~al. (\cite{Nak06}), who proposed that the dark
GRB 051022 took place inside a dusty molecular cloud, given the large
extinction ($A_{\rm V} = 49$~mag) derived from X-ray data. We point out that a
very high dust extinction has been also proposed as a possible reason for the
discrepancy between the rate of core-collapse SNe (which are related to LGRBs)
determined from NIR surveys, and that estimated from the FIR luminosity of
their parent galaxies (Mannucci et~al. \cite{Man03}). However, we note
that the only LGRB with a HG possibly showing a similar extinction ($A_{\rm V}
= 3.4$~mag according to Vreeswijk et~al. \cite{Vre01}, but only 0.15~mag
according to Christensen et~al. \cite{Chr04}) to that of our candidate, is not
a DGRB (GRB 990712 at $z = 0.43$). If its high extinction is
confirmed, this could be the result of a different dust distribution in both
galaxies, obscuration not affecting the star-forming regions in the HG of
GRB 990712 (not an e(a) type). Or, this might imply that dust
extinction alone is not a sufficient condition to produce a DGRB. On the other
hand, the confirmation of a low extinction for the GRB 990712 HG would
strengthen the hypothesis that dusty strabursts produce DGRBs. It is important
to note, however, that there are DGRB HGs which are not representative of dusty
starbursts (e.g., GRB 000210; Gorosabel et~al. \cite{Gor03}).

Our measurement of a redshift $z = 0.584$ for the likely host galaxy of
GRB 050223 casts a shadow over the high redshift hypothesis proposed
by Page et~al. (\cite{Pag05}) to explain the faintness of its X-ray afterglow,
leaving the possibility of a large jet opening or viewing angle. Our detection
of obscured star formation in its HG does not contradict their conclusion of
the GRB 050223 afterglow being underluminous. Intrinsic
faintness
might well be complementary to dust absorption in producing a DGRB, either by
requiring a modest dust mass to suppress the afterglow, or by being ineffective
in destroying the circumburst dust. We note that GRB 050223 and
GRB 990712 lie at opposite ends in the X-ray afterglow brightness
\-distribution.

Although it shows a high optical extinction, the likely HG of GRB
050223 does not seem to be one of the galaxies with luminous, reddened IR
starbursts and very high SF activity. Particularly, it was not detected by {\em
Spitzer} (Le Floc'h 2006, priv. comm.) in the FIR. The constraint set by these
observations, ${\rm SFR} < 10$~$M_{\sun}$~yr$^{-1}$, is consistent with our
results. However, it is noteworthy that the position of this HG in Figs.~4 and
5 of Le Floc'h et~al. (\cite{LeF03}) is consistent with the loci of the NIR
counterparts of high redshift ISO galaxies. Further investigation of this HG
would be important to get a deeper insight into the nature of DGRBs and the
LGRB-SF connection.

\begin{acknowledgements}

We acknowledge D.~Malesani for a careful reading of the manuscript and useful
comments on it, and the anonymous referee for suggestions that greatly improved
this work. The MISTICI activities in Italy are supported by ASI grant
I/R/039/04. This publication makes use of data products from the Two Micron All
Sky Survey, which is a joint project of the University of Massachusetts and the
Infrared Processing and Analysis Center / California Institute of Technology,
funded by the National Aeronautics and Space Administration and the National
Science Foundation.

\end{acknowledgements}


\begin{thebibliography}{}

\bibitem[2003]{Ber03}
Berger, E., Cowie, L.~L., Kulkarni, S.~R., et~al. 2003,
\apj, 588, 99

\bibitem[2006]{Ber06}
Berger, E., \& Shin, M.-S. 2006,
GCN 5283

\bibitem[2005]{Blu05}
Blustin, A., Branduardi-Raymont, G., Breeveld, A., et~al. 2005,
GCN 3093

\bibitem[2006]{Cam06}
Campana, S., Mangano, V., Blustin, A.~J., et~al. 2006,
Nature, 442, 1008

\bibitem[2004]{Chr04}
Christensen, L., Hjorth, J., Gorosabel, J., et~al. 2004,
\aap, 413, 121

\bibitem[2005]{DeL05}
De Luca, A., \& Campana, S. 2005,
GCN 3109

\bibitem[1990]{Dic90}
Dickey, J.~M., \& Lockman, F.~J. 1990,
\araa, 28, 215

\bibitem[2001]{Djo01}
Djorgovski, S.~G., Frail, D.~A.; Kulkarni, S., et al. 2001,
ApJ, 562, 654

\bibitem[2002]{Elb02}
Elbaz, D., Cesarsky, C.~J., Chanial, P., et~al. 2002,
\aap, 384, 848

\bibitem[1999]{Fry99}
Fryer, Ch., Woosley, S.~E., \& Hartmann, D.~H. 1999,
\aap, 526, 152

\bibitem[2005]{Gio05}
Giommi, P., Capalbi, M., Perri, M., et~al. 2005,
GCN 3054

\bibitem[2003]{Gor03}
Gorosabel, J., Christensen, L., Hjorth, J., et~al. 2003,
\aap, 400, 127

\bibitem[2005]{Gor05}
Gorosabel, J., Eguchi, S., de Ugarte Postigo, A., et~al. 2005,
GCN 3061

\bibitem[2005]{Gro05}
Gronwall, C., Blustin, A., Brown, P., et~al. 2005,
GCN 3057

\bibitem[1998]{Gro98}
Groot, P.~J., Galama, T.~J., van Paradijs, J., et~al. 1998,
\apj, 493, L27

\bibitem[2006]{Hai06}
Haislip, J.~B., Nysewander, M.~C., Reichart, D.~E., et~al. 2006,
Nature, 440, 181

\bibitem[2003]{Hjo03}
Hjorth, J., Sollerman, J., Moller, P., et al. 2003,
Nature, 423, 847

\bibitem[2004]{Jak04}
Jakobsson, P., Hjorth, J., Fynbo, J.~P.~U., et al. 2004,
\apj, 617, L21

\bibitem[2005]{Jak05}
Jakobsson, P., Frail, D.~A., Fox, D.~B, et al. 2005,
\apj, 629, 45

\bibitem[2006]{Kaw06}
Kawai, N., Kosugi, G., Aoki, K., et~al. 2006,
Nature, 440, 184

\bibitem[1998]{Ken98}
Kennicutt, R.~J. 1998,
\araa, 36, 189

\bibitem[2002]{LeF02}
Le Floc'h, E., Duc, P.-A., Mirabel, I.~F., et~al. 2002,
\apj, 581, L81

\bibitem[2003]{LeF03}
Le Floc'h, E., Duc, P.-A., Mirabel, I.~F., et~al. 2003,
\aap, 400, 499

\bibitem[2006]{LeF06}
Le Floc'h, E., Charmandaris, V.; Forrest, W., et al. 2006,
\apj,~642,~636

\bibitem[2006]{Lew06}
Levan, A., Fruchter, A., Rhoads, J., et~al. 2006,
\apj,~647,~471

\bibitem[2003]{Lis03}
Liske, J., Lemon, D.~J., Driver, S.~P., et~al. 2003,
\mnras, 344, 307.

\bibitem[2005]{Mer05}
Mereghetti, S., Gotz, D., Mowlavi, N., et~al. 2005,
GCN 3059

\bibitem[2004]{Mal04}
Malesani, D., Tagliaferri, G., Chincarini, G., et~al. 2004,
\apj, 609, L5

\bibitem[2003]{Man03}
Manucci, F., Maiolino, R., Cresci, G., et~al. 2003,
\aap, 401, 51

\bibitem[2005]{Mit05}
Mitani, T., Barbier, L., Barthelmy, S., et~al. 2005,
GCN 3055

\bibitem[2006]{Mor06}
Moretti, A., Perri, M., Capalbi, M., et al. 2006,
\aap, 448, L9

\bibitem[2006]{Nak06}
Nakagawa, Y.~E., Yoshida, A., Sugita, S., et al. 2006,
PASJ, 58, L35

\bibitem[2005]{Nys05}
Nysewander, M., Bayliss, M., Haislip, J., et~al. 2005,
GCN 3067

\bibitem[2005]{Pag05}
Page, K.~L.; Rol, E.; Levan, A.~J., et al. 2005,
\mnras, 363, L76

\bibitem[2006]{Ped06}
Pedersen, K., Hurley, K., Hjorth, J., et al. 2006,
\apj, 636, 381

\bibitem[2002]{Pir02}
Piro, L., Frail, D.~A., Gorosabel, J., et al. 2002,
\apj, 577, 680

\bibitem[1997]{Pog97}
Poggianti, B.~M. 1997,
\aap~Suppl.~Ser., 122, 399

\bibitem[2000]{Pog00}
Poggianti, B.~M., \& Wu, H. 2000,
\apj, 529, 157

\bibitem[2001]{Pog01}
Poggianti, B.~M., Bressan, A., \& Franceschini, A. 2001,
\apj, 550, 195

\bibitem[2004]{Pro04}
Prochaska, J.~X., Bloom, J.~S., Chen, H.-W., et~al. 2004,
\apj, 611, 200

\bibitem[2006]{Sar06}
Saracco, P., Fiano, A., Chincarini, G., et~al. 2006,
\mnras, 367, 349

\bibitem[1998]{Sch98}
Schlegel, D.~J., Finkbeiner, D.~P., \& Davis, M. 1998,
\apj, 500, 525

\bibitem[2005]{Smi05}
Smith, D.~A. 2005,
GCN 3056

\bibitem[2003]{Sta03}
Stanek, K.~Z., Matheson, T., Garnavich, P.~M., et~al. 2003,
\apj, 591, L17

\bibitem[2005]{Tag05}
Tagliaferri, G., Antonelli, L., Chincarini, G., et~al. 2005,
\apj, 443, L1

\bibitem[2001]{Vre01}
Vreeswijk, P.~M., Fruchter, A., Kaper, L., et.~al. 2001,
\apj, 546, 672

\bibitem[1991]{Win91}
Windhorst, R., Burstein, D., Mathis, D.~F., et.~al. 1991,
\apj, 380, 362

\end{thebibliography}
\end{document}